\documentclass[aps,prd,singlecolumn,groupedaddress,showpacs,preprintnumbers,nofootinbib,draft]{revtex4}
\usepackage[dvipdfm]{graphicx}
\usepackage{amssymb}
\usepackage{amsmath}
\usepackage{parskip}
\usepackage{epsfig}
\usepackage{dcolumn}
\usepackage{rotating}
\usepackage{graphpap}
\newcommand{\bea}{\begin{eqnarray}}
\newcommand{\eea}{\end{eqnarray}}
\newcommand{\be}{\begin{equation}}
\newcommand{\ee}{\end{equation}}
\newcommand{\beqy}{\begin{eqnarray}}
\newcommand{\eeqy}{\end{eqnarray}}

\newcommand{\mx}{\mbox}
\newcommand{\mt}{\mathtt}

\newcommand{\bb}{\beta}

\newcommand{\Ga}{\Gamma}

\newcommand{\de}{\delta}

\newcommand{\Om}{\Omega}

\newcommand{\cO}{{\cal O}}

\newcommand{\ra}{\rightarrow}
\newcommand{\Ra}{\Rightarrow}

\newcommand{\LF}{\left(}
\newcommand{\RF}{\right)}
\newcommand{\LT}{\left[}
\newcommand{\RT}{\right]}
\newcommand{\mtx}{\mt{max}}
\newcommand{\ie}{{\it i.e.\ }}

\begin{document}
\title{Emergence of a Cyclic Universe from the Hagedorn Soup}
\vspace{3cm}
\author{Tirthabir Biswas}
\vskip 3cm
\affiliation{
{\it e-mail: tbiswas@gravity.psu.edu\\
Department of Physics,\\
Institute for Gravitation and the Cosmos,\\
The Pennsylvania State University,\\
104 Davey Lab, University Park, PA,16802, U.S.A \\}
\vskip 1mm
}


\date{\today}

\begin{abstract}
One of the challenges of constructing a successful cyclic universe scenario is  to be able to incorporate the second law of thermodynamics which typically leads to Tolman's problem of ever shrinking cycles. In this paper we construct a non-singular toy model where as the cycles shrink in the past they also spend more and more time in the entropy conserving Hagedorn phase. Thus in such a scenario the entropy asymptotes to a finite non-zero constant in the infinite past. The universe ``emerges'' from a small (string size) geodesically complete  quasi-periodic space-time. This  paradigm also naturally addresses some of the classic puzzles of Big Bang cosmology, such as the largeness, horizon and flatness problems.
\end{abstract}


\maketitle

\newpage

{\it Introduction:}
Cyclic models of universe offer a promising  resolution to the Big Bang singularity problem~\cite{tolman,narlikar,ekcyclic,barrow,review}. The hope is that rather than having to deal with a ``beginning of time'' when the universe starts from a singularity, time can be made ``eternal'' in either direction (past and future). However, there are three main reasons why it still remains an extreamely challenging task to construct a theoretically consistent and phenomenologically viable model: {\it (i)} In General Relativity cyclic universes involve violation of various energy conditions at the ``bounce''~\cite{paris-ec} where contraction gives way to expansion and typically this requires invoking problematic matter sources, such as ghosts. {(\it ii)} We are confronted with the second law of thermodynamics according to which the total entropy in the universe can only increase monotonically. As first pointed out by Tolman~\cite{tolman}, this immediately tells us that the evolution can at most be ``quasi-periodic'' where the length of the cycles monotonically increase with the increase in entropy from cycle to cycle. Moreover,  the problem of the ``beginning of time'' comes back to haunt us, as  the cycles  become vanishingly small at a finite proper time  in the past~\cite{tolman}. {\it (iii)}  One of the biggest successes of inflationary cosmology  is, that it not only explains why the universe is so large, flat and homogeneous today, but also provides a causal mechanism to produce small inhomogeneities which we observe in CMB and LSS. Thus any seriously competing  alternative to the standard inflationary paradigm should be able to  reproduce these basic successes of inflation.

The main purpose of this paper is to address in a theoretically consistent (ghost and singularity free) framework {\it (i)}, Tolman's problem of ever-shrinking cycles ({\it ii}) by taking recourse to a crucial feature of string theory,  the existence of the Hagedorn phase where all the different string states (massless and massive) are in thermal equilibrium and therefore entropy is conserved. This allows us to construct an ``emerging cyclic universe'' where as one goes back in cycles (time) and the cycles become shorter and on an average hotter, it spends more and more time in the Hagedorn phase where no entropy is produced. Thus the universe asymptotes to a constant entropy state with almost-periodic contractions and expansions. One of the key difference between the approaches of this paper to previous ones is that we incorporate  non-singular bounces where the Hubble rate remains  bounded.  In a singular big crunch/bang scenario the interaction rates which can potentially maintain thermal equilibrium among the different  species,  cannot keep up with the {\it diverging}  Hubble rate.   Thus most previous literature  assumes that such transitions will produce large amounts of entropy which eventually leads to Tolman's  problem.

In the model we propose, entropy is produced afterwards  when the different species in the Hagedorn phase fall out of thermal equilibrium below some critical  temperature, say $T_p$, and energy starts to flow from hotter to colder species.  Consequently, we find a class of solutions where the length of the cycles increase monotonically with time. Each of these cycles spend some amount of time in both the Hagedorn and non-Hagedornic phase, and as a result some amount of entropy is always produced.  However, as one goes back in the past the universe spends less and less time in the entropy producing non-Hagedornic phase, the entropy produced in a given cycle therefore goes to zero, and  the universe asymptotes towards a periodic evolution.

Apart from ({\it i}) and ({\it ii}), this new cosmological scenario naturally explains the flatness, largeness  and horizon problems. The universe can ``start'' out small ($\sim$ string length) with a low entropy and with curvature density around the string scale and yet through the course of entropy production  in the infinite cycles in the past, the evolution eventually produces cycles with a large entropy  where curvature remains negligible for a long time.  Through the course of these infinite number of cycles there is also obviously enough time to establish causal connection.  The only {\it a priori} assumption that we have to make is that of spatial homogeneity. Thus an  important open question  is whether the ``initial cycles''  can drive one to homogeneity? Finally, we do not attempt to address ({\it iii}) in this paper, but we point to mechanisms that have been invoked~\cite{ekcyclic,robert} to produce nearly scale-invariant spectrum without inflation, and that encouraging progress have been made to include these in the framework of non-singular bouncing cosmologies~\cite{abb,bbmw}.

\noindent{\it Hagedorn Physics, Casimir Energy and Bounces: }
We start by considering the Hagedorn phase of matter on a space-time which includes our observed three dimensional universe, along with appropriately compactified extra dimensions. For us, the most crucial property  of the Hagedorn phase  is, that in this phase all the different string states are in thermal equilibrium with each other and the entropy  remains conserved. In our analysis we will only care about the ``universal'' leading order behaviour of entropy~\cite{jain,vafa-hagedorn}
\be
S=\bb_H E+\cO(VT_H^4/E)\Ra \rho_{\mt{hag}}={T_H^4S/V}
\label{hag-entropy}
\ee
where $S$ and $E$ are the entropy and energy of the Hagedorn phase, $V$ is the spatial volume of the universe, and $T_H$ is the Hagedorn temperature.

As mentioned before, finding bounces with consistent physics is challenging, but the existence of negative Casimir energies offers an interesting possibility. As a simple illustrative example, in this paper we will consider Casimir energies coming from minimally coupled free scalar fields\footnote{In string theory massless moduli scalar fields are ubiquitous, but typically they also couple to the Ricci scalar. Such couplings may modify the form of the casimir energy. Also, these scalars can be dynamic in the early universe which will modify the evolution, but these questions can hopefully be answered through a more elaborate study in the future.} in a closed universe setting. It was shown in~\cite{casimir} that in this case the Casimir energy $\sim  -a^{-4}(t)$. The Hubble equation  therefore reads as
\be
H^2={T_H^4\over 3 M_p^2}\LT {S\over a^3}-{\Omega_c\over a^4}-{\Omega_k\over a^2}\RT
\label{hag-hubble}
\ee
where we have defined a dimensionless ``scale factor'' $a\equiv T_H V^{1/3}$ and  $\Om_c,\Om_k$ are dimensionless $\sim\cO(1)$ constants associated with the Casimir energy~\cite{casimir}, and spatial curvature respectively.

Now, if  both the bounce and the turnaround occurs within the Hagedorn phase then we end up having an ``eternally periodic cyclic universe'' given as a function of the conformal time, $\tau$:
\be
a(\tau) = {S-\sqrt{S^2-4\Om_c\Om_k}\cos \nu\tau\over 2\Om_k}\mx{ with } \nu\equiv \sqrt{\Om_k\over 3}{T_H^2\over M_p}
\label{periodic}
\ee
Phenomenologically this is rather uninteresting, as the temperature  cannot fall below $T_p$ which is expected to be close to the string scale. We are therefore interested in the quasi-periodic evolution where cycles can grow with the production of entropy, once out of the Hagedorn phase.

\noindent{\it Ordinary matter, Curvature and Turnarounds: }
From simple thermodynamic considerations it follows that once the different species fall out of equilibrium and there is exchange of energy (from the hotter to the colder species), entropy in generated.  We want to incorporate such entropy production  in a toy model setting. For this purpose we will consider a two species model with radiation, $\rho_r$, and non-relativistic matter,  $\rho_m$, and assume that the entire string gas can be clubbed into  two of these categories near the transition from Hagedorn phase to radiation. Thus the picture is, that both these species are in thermal equilibrium in the Hagedorn phase, but after the transition temperature, the two species fall out of equilibrium and energy starts to flow from the dust-like matter to radiative degrees of freedom.

Before we model this energy exchange, it is instructive to look at the usual solution of a closed universe with ordinary matter and radiation. The Hubble equation and it's solution is of the same form as (\ref{hag-hubble}) and (\ref{periodic})  respectively with the substitution, $S\ra\Om_m$ and $\Om_c\ra \Om_c-\Om_r$.

To complete the story one needs to relate the quantities in the two halves of the cycle.  To keep the calculations simple, we are  going to assume that one can ignore $\Om_c$ as compared to $\Om_r$ which is always possible provided $\Om_c\ll\Om_k^2$~\cite{companion}. Now, unfortunately we do not have access to the complicated transitionary dynamics from the Hagedorn to the radiation dominated epoch as the corrections in (\ref{hag-entropy}) are not under control, but by our explicit construction the total entropy should be conserved till the transition point, leading to
\be
S=b_r \Om_r^{3/4}a_p^3+b_m \Om_m a_p^3
\ee
where the constants $b_r,b_m$ depend on the number of radiative and non-relativistic species one has, and their properties. We also know that at the point of phase transition the matter and radiation were still just in thermal equilibrium, \ie had the same temperature
\be
T_m={T_H\over b_m}={T_H4\Om_{r}^{1/4}\over  3b_ra}=T_r
\ee
Finally, we make a phenomenological ansatz about the relative energy densities of matter and radiation at the transition epoch
\be
\mu\equiv \rho_{m}/\rho_r=\Om_{m}a_p/\Om_r
\ee
Ideally, this ratio should be calculable if we understood the transition from Hagedorn to radiation+matter phase, but phenomenological considerations (matter-radiation equality  $\sim 100\ ev$) already suggests $\mu$ to be very small $\sim 10^{-22}$.

We can now determine all the dynamical quantities involved in the second half in terms of the  entropy of the Hagedorn phase (which varies from cycle to cycle), and phenomenological constants, $b_m,b_r$ and $\mu$~\cite{companion}. Under the simplifying assumptions $b_m\sim 1$ and $\mu,\Om_c/\Om_k^2\ll 1$ we have
\be
\Om_r=\LF{S\over b_r}\RF^{4/3}\ ,\ \Om_m={3\over4}\mu S\mx{ and }a_p={4S^{1/3}\over 3b_r^{4/3}}
\label{matching}
\ee

\noindent{\it Entropy Production:}
Having understood the behaviour of our universe in a ``non-interacting'' entropy-preserving setting, let us now try to understand how energy exchange between ordinary matter and radiation effects the dynamics.
The energy exchange process can be captured via phenomenological equations of the form~\cite{barrow}
\be
\dot{\rho}_r+4H\rho_r=T_H^4 s\mx{ and }\dot{\rho}_m+3H\rho_m=-T_H^4s
\label{continuity}
\ee
which preserves conservation of stress-energy tensor, but which ``typically'' breaks the time reversal symmetry  providing an ``arrow of time''. $s$ characterizes the energy flow and one expects it to depend on the densities and the scale factor.  It is easy to compute the net entropy production in such models:
\be
\dot{S}=\dot{S}_r+\dot{S}_m=a^3s\LF{3b_rT_H\over 4\rho_r^{1/4}}-b_m\RF=a^3s\LF{T_H\over T_r}-{T_H\over T_m}\RF
\label{entropy-growth}
\ee
We will assume $s>0$. Consistency with $2^{nd}$ law of thermodynamics then implies that the quantity within brackets must be positive. This is nothing but the condition that the temperature of the non-relativistic gas be greater than the radiative gas, so that energy flows from the hotter non-relativistic species to colder radiation in accordance with $1^{st}$ law of thermodynamics. Since in our picture the two species have the same temperature $T_p$, at the transition point, where after $T_r$ decreases, while $T_m$ stays fixed, (\ref{entropy-growth}) is consistent with both the $1^{st}$ and  $2^{nd}$ law of thermodynamics.

Let us now see why we can realize the emergent cyclic universe. It is sufficient to look at small cycles which (we claim) asymptotically approaches the periodic evolution (\ref{periodic}). In these cycles matter density is always negligible as compared to radiation, so that the expression for the turnaround point reduces to
\be
a_{\mtx}\approx {\sqrt{\Om_r/\Om_k}}\approx S^{2/3}/(  \sqrt{\Om_k}b_r^{2/3})
\ee
 We notice that $a_{\mtx}\sim S^{2/3}$,  while $a_p\sim S^{1/3}$. In other words as we go back in the past and the entropy decreases, $a_p$ catches up with $a_{\mtx}$, and the universe spends less and less time in this entropy-generating phase. In turn, less and less entropy is produced,  and in fact, the entropy approaches a constant given by
\be
a_{\mtx}=a_p\Ra \lim_{n\ra -\infty}S_n= S_{cr}\equiv (4/ 3)^3\Om_k^{3/2}b_r^{2}
\label{critical}
\ee

It is not difficult to estimate  the entropy growth for ``small'' cycles defined by $\nu\tau_p\ll 1$, where $\tau_p$ is the conformal time corresponding to the transition point $a_p=a(\tau_p)$. Under the technically simplifying assumption   $s\ll M_p/T_H^2$,   one finds by solving (\ref{entropy-growth}), the increase in entropy in a given cycle to be given by~\cite{companion}
\be
\Delta S={2a_{\mtx}^4s_{cr}\nu^2\tau_p^3\over 3a_p}\approx {d\de S\over dn}\mx{ where }s_{cr}\equiv s(a_p(S_{cr}))
\label{deltaS}
\ee
where ``$n$'' labels the number of the cycle and $\de S\equiv S-S_{cr}$.
 Since, $a_{\mtx}, a_p$ and $\tau_p$ are all known functions of the entropy, this differential equation can be solved.
As $n\ra -\infty$ we find the leading order result:
\be
S\approx S_{cr}\LT1+{1\over C^2 n^2}\RT\mx{ where } C= 8\LF{2\over 3}\RF^{11/2}{s_{cr}\over \nu b_r^4}
\ee
We now explicitly see that, $S\ra S_{cr}$ at the past infinity.

We note in passing that the divergence of entropy at $n=0$ is only an artifact of approximating $s$ by a constant in deriving (\ref{deltaS}), which breaks down once the cycles become large. For large cycles the growth of entropy with $n$ would depend on the modeling of $s$ (see~\cite{companion} for an illustrative example), but on physical grounds we expect $S_n$ to diverge (if at all) only as $n\ra\infty$.

\noindent{\it Conclusions:}
We have proposed a new non-singular cyclic cosmology which addresses Tolman's problem of ever shrinking cycles as well as some of the standard puzzles of Big Bang cosmology, such as the horizon, entropy, largeness, and flatness problems.
The success of the scenario largely rests on the crucial assumption of the existence of a thermal stringy Hagedorn phase near the bounce.  One expects this to hold if the string interaction rate $\Ga_{\mt{str}}\gg H_{\mt{max}}$. Based on naive  estimates of $\Ga_{\mt{str}}$~\cite{anupam} one can check that this would be true as long as $e^{2\psi}>T_H/100 M_p$ \cite{companion}, where $\psi$ is the dimensionally reduced dilaton~\cite{anupam}. This seems a rather reasonable bound to satisfy, however one would like to perform a more careful investigation  in the future.

Likewise, to make this scenario more concrete we need to include moduli dynamics/stabilization (which should also explain why the observed three dimensions became large as compared to the extra dimensions, see~\cite{vafa} for a possible relevant mechanism), sources of stringy Casimir energies and mechanisms for generating near-scale invariant perturbations,  but at the least the scenario looks promising.

\end{document}